# Sustainable bioplastics from amyloid fibril-biodegradable polymer blends

*Mohammad Peydayesh, Massimo Bagnani, and Raffaele Mezzenga\**

ETH Zurich, Department of Health Sciences and Technology, Schmelzbergstrasse 9, Zurich 8092, Switzerland

E-mail: raffaele.mezzenga@hest.ethz.ch

Plastic waste production is a global challenging problem since its accumulation in the environment is causing devastating effects on the planet's ecosystem. Sustainable and green solutions are urgently needed, and this pairs with increasingly stronger regulations combined with improved ecological awareness. This study proposes a simple, scalable and water-based process to produce free-standing, transparent and flexible bioplastic films by combining amyloid fibrils with biodegradable polymers as two main building blocks. Amyloid fibrils can be obtained through denaturation and self-assembly from a broad class of food proteins found in milk, soy, and egg, for example. Whey is used here as a model protein, since it is the major by-product of dairy industries, and its valorization creates a valuable opportunity to produce sustainable, biodegradable, and environmentally friendly bioplastics perfectly integrated within a circular economy. Against this background, we highlight the sustainability superiority of these bioplastics over common plastics and bioplastic via a detailed life cycle assessment, anticipating an important role of this new class of bioplastics in mitigating the pressing plastic pollution challenge.

## 1. Introduction

Plastic is one of the most abundant man-made materials and, although its widespread use started only 70 years ago, an estimate of 8300 million metric tons (Mt) of plastic have been produced to date. The intensive utilization of single-use containers drastically accelerated plastic production, and packaging is now plastic's largest market[1]. Approximately 150 Mt of solid plastic, corresponding to half of the annual global production is thrown away each year worldwide[2]. Most polymers used to produce plastics are derived from fossil hydrocarbons, posing a threat to fossil fuel sources. Additionally, most common plastics are not biodegradable and, if not permanently destroyed by proper thermal treatment, accumulate in the natural environment or landfills causing devastating effects to the planet's ecosystem. Up to 80% of the global plastic waste ends up contaminating the environment, and 4-12 Mt of plastic enters the oceans each year[3]. Plastic debris can be found in all the ocean basins[4] and are so abundant in the environment that it can be used as a geological indicator of the Anthropocene era[5]. Although recycling is suggested as a possible solution to mitigate some of these problems, this process remains limited to less than 9% of global plastic waste since it is costly, time-consuming, cannot be applied to many polymeric materials, and the quality of the polymers obtained is low[2].

Thanks to the growing awareness about the environmental issues related to the accumulation and disposal of traditional plastics, stringent regulations are being implemented worldwide, pushing the plastic market towards a transition to more sustainable products and processes. The production of bioplastic increased dramatically in the past few years, and it is expected to grow even more substantially in the future[6–8]. The European Bioplastic organization categorizes bioplastic into two major groups defined as biodegradable and bio-based plastics[9]. The latter are produced employing renewable resources such as cellulose, starch polylactide (PLA), or polyhydroxyalkanoates (PHAs) instead of fossil fuels; however, unfortunately, not all bio-based plastics are biodegradable, such as those produced by converting renewable resources like corn, sugarcane, and cassava into building blocks for polyethylene terephthalate (PET)[9,10]. The American Society for Testing and Materials (ASTM) defines biodegradable those plastics that degrade under the action of naturally occurring microorganisms such as bacteria, fungi, and algae[11]. Additionally, if a plastic degrades due to biological processes into biomass, carbon dioxide, water, and inorganic compounds without leaving toxic residues, it is defined as compostable. So all compostable plastics are biodegradable; however, not all biodegradable plastics are compostable[11].

Protein-based bioplastic is attracting tremendous attention due to its broad availability, fast biodegradation rates, and food-grade nature resulting in films that can even be classified as edible[12–15]. The main drawbacks of protein-based bioplastic derive from the intrinsic nature of native protein monomers, which are often globular, hydrophilic, and water-soluble and result in difficulties to process films, show poor mechanical and barrier properties, and are very sensitive to water and humidity.

In the dairy industries, for producing 1 kg of cheese, 8-9 kg of whey are produced as a by-product. It represents the main by-product of the dairy industries, where each year, approximately 120 million tons of whey are produced globally. Since only half of the whey produced is transformed into valuable products such as human or animal feed[16] (**Figure 1**), the disposal of surplus whey represents a crucial issue for the dairy industries and causes environmental concerns due to the high biological oxygen demand by-product[17–19]. In fact, whey contains a high load of organic matter, which is mainly composed of lactose (0.18-60 kg/m$^3$), proteins (1.4-33.5 kg/m$^3$), and fats (0.08-10.58 kg/m$^3$)[18,20]. Whey is, therefore, a rich source of proteins whose the most abundant is β-lactoglobulin[21,22], a globular protein that can easily self-assembly into amyloid fibrils[23–25].

Amyloid fibrils play critical functional roles in various biological processes in multiple organisms, ranging from bacteria to humans. Thanks to their promising biophysical properties, mechanical and chemical stability, many applications have been proposed[26–29]. Amyloid fibrils can also self-assembly *in vitro* from various proteins in milk, egg, and soy through denaturation and hydrolysis under proper conditions, typically involving low pH and high temperatures[30–32]. A broad range of functionalities characterizes the fibrils obtained from these food proteins and, thanks to their remarkable properties that are far superior to those of single monomers, such as high stiffness and aspect ratios, they have been used as building blocks for developing suspensions, emulsions, membranes, and gels with high performances[28,29,32,33].

This study focuses on developing bioplastic composed of amyloid fibrils and showcase them as ideal candidates to produce hybrid films. In particular, we show that amyloid fibrils can be used as building blocks for engineering novel bioplastics with targeted characteristics, further tuned by blending different functional additives such as bio-polymers and plasticizers to improve the performances of the resulting films. Other important film properties such as water stability, hydrophobicity, and antioxidant activity can be tuned by chemical treatment or coatings. These novel bioplastics are characterized by a wide range of properties that can be achieved and tuned without requiring the use of non-biodegradable or toxic compounds. Moreover, these bioplastics show great potential for commercialization with economic viability thanks to the meager cost of the protein used, which are mainly obtained by waste products of the food industry, and the cheap, scalable, and environmentally friendly water-based production protocol for film formation. Ultimately, producing bioplastic using food wastes helps industry in two ways: not only it improves their production in terms of sustainability, waste management, and valorization, but it also directly improves their circular economy.

## 2. Results and disscussion

### 2.1. Bioplastic films

The bioplastic presented here can be obtained by a simple, scalable, and water-based protocol which is schematically described in Figure 1. Specifically, WPI (whey protein isolate) -here taken as a model protein available from food processing byproducts- is dispersed in pH 2 water together with a plasticizer (Glycerol) and a water-soluble polymer (such as poly (vinyl alcohol), PVA). The solution is then heated at 90°C for 5 h to solubilize all the components homogeneously and convert whey protein (mostly β-lactoglobulin) into self-assembled amyloid fibrils. Free-standing homogeneous and transparent films (see **Figure 1**) can be obtained by casting the solution onto suitable substrates followed by solved evaporation.

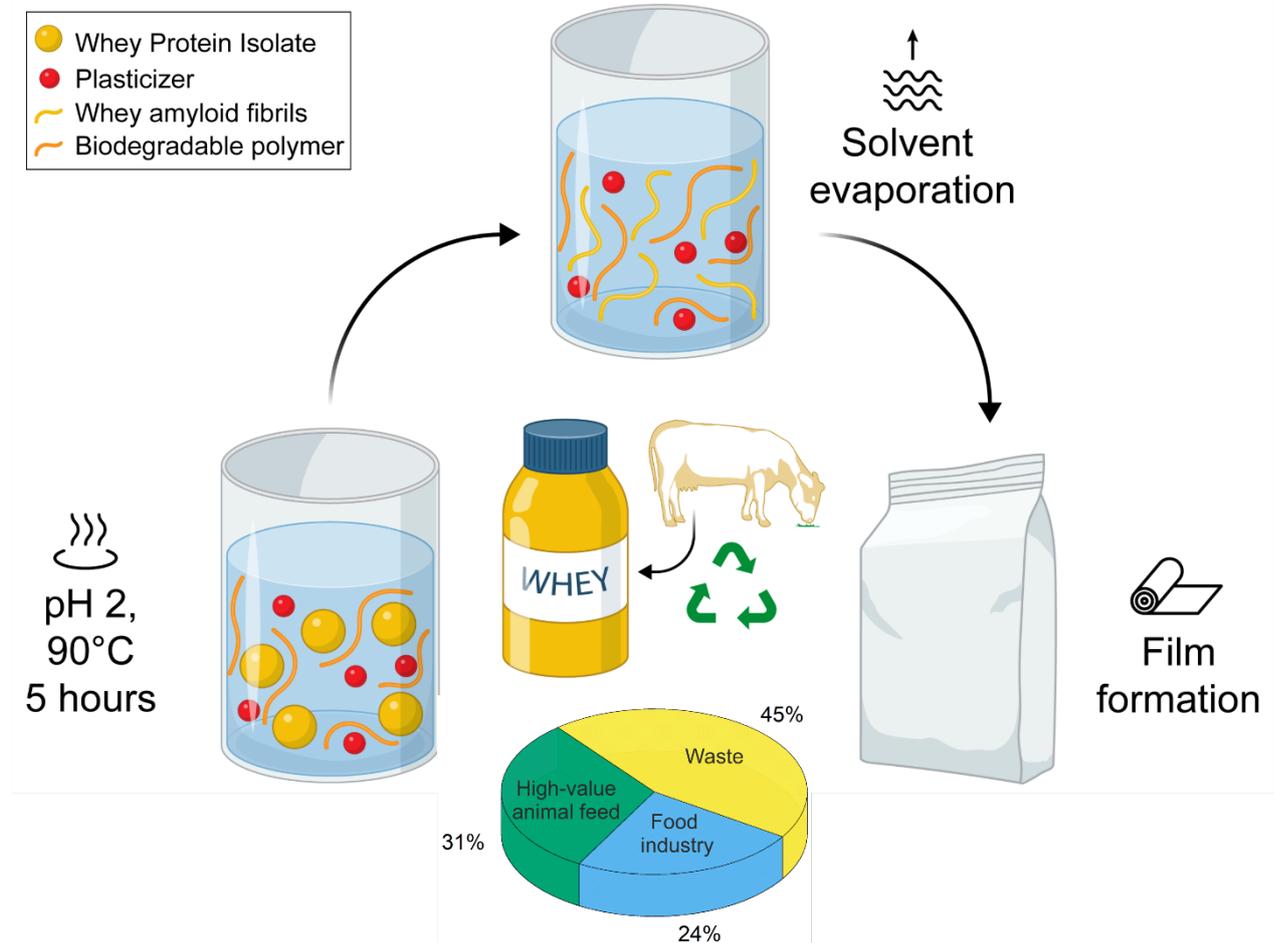

**Figure 1.** Schematic representation of the bioplastic preparation protocol. Each year in Switzerland 1.3 Mt of whey are produced, of which only half of it is valorized efficiently[16]. Whey protein isolate, Glycerol, and PVA are dispersed in water, the solution is heated at 90°C for 5 hours to allow β- lactoglobulin monomers to self-assemble into amyloid fibrils. The solution is then cast to obtain bioplastic film through solvent evaporation.

## 2.2. Microstructure

To verify that amyloid fibrils self-assembly is not inhibited by the other compounds in solution (Glycerol and PVA and other compounds contained in WPI), atomic force microscopy (AFM) has been used to image the solution resulting after the heat treatment. **Figure 2a** shows amyloid fibrils obtained from pure β-lactoglobulin dispersed in pH 2 water. As a comparison, **Figure**

**2b** shows the amyloid fibrils formed from WPI and in the presence of PVA and Glycerol, confirming that the other compounds do not inhibit the self-assembly of β-lactoglobulin into amyloid fibrils in solution. The main difference between the fibrils obtained appears to be their contour length distribution that decreases when amyloids are produced in the other compounds' presence. This difference derives from the fact that a very high viscosity characterizes the solution used to produce bioplastic. To avoid gelation, vigorous stirring has to be applied during the heat treatment. In fact, these mechanical stresses are known to induce shortening of amyloid aggregates[34].

In **Figure 2 c** and **d**, the SEM images of the surface of the hybrid amyloid and monomer films are presented. As shown in the figure, the amyloid-based films' surface is smooth, homogenous, and without cracks. However, the surface of the films obtained with WPI monomers shows multiple cracks spanning several micrometers. These cracks increase the oxygen and vapor transport through the films, resulting in films that might not be suitable for food packaging. Whey has already been proposed as the right candidate to obtain edible films[35–37], thanks to the low cost and broad availability of this by-product. However, the film resulting from the whey protein in their native or partially hydrolyzed state results in low mechanical properties and water stability [35–37].

On the contrary, amyloid based bioplastic results in a film characterized by a highly homogeneous surface, thanks to the high degree of interactions between amyloid fibrils, the plasticizer (Glycerol), and the polymer chains (PVA), favored by the very high aspect ratio of amyloid fibrils combined with the numerous functional groups on their surface.

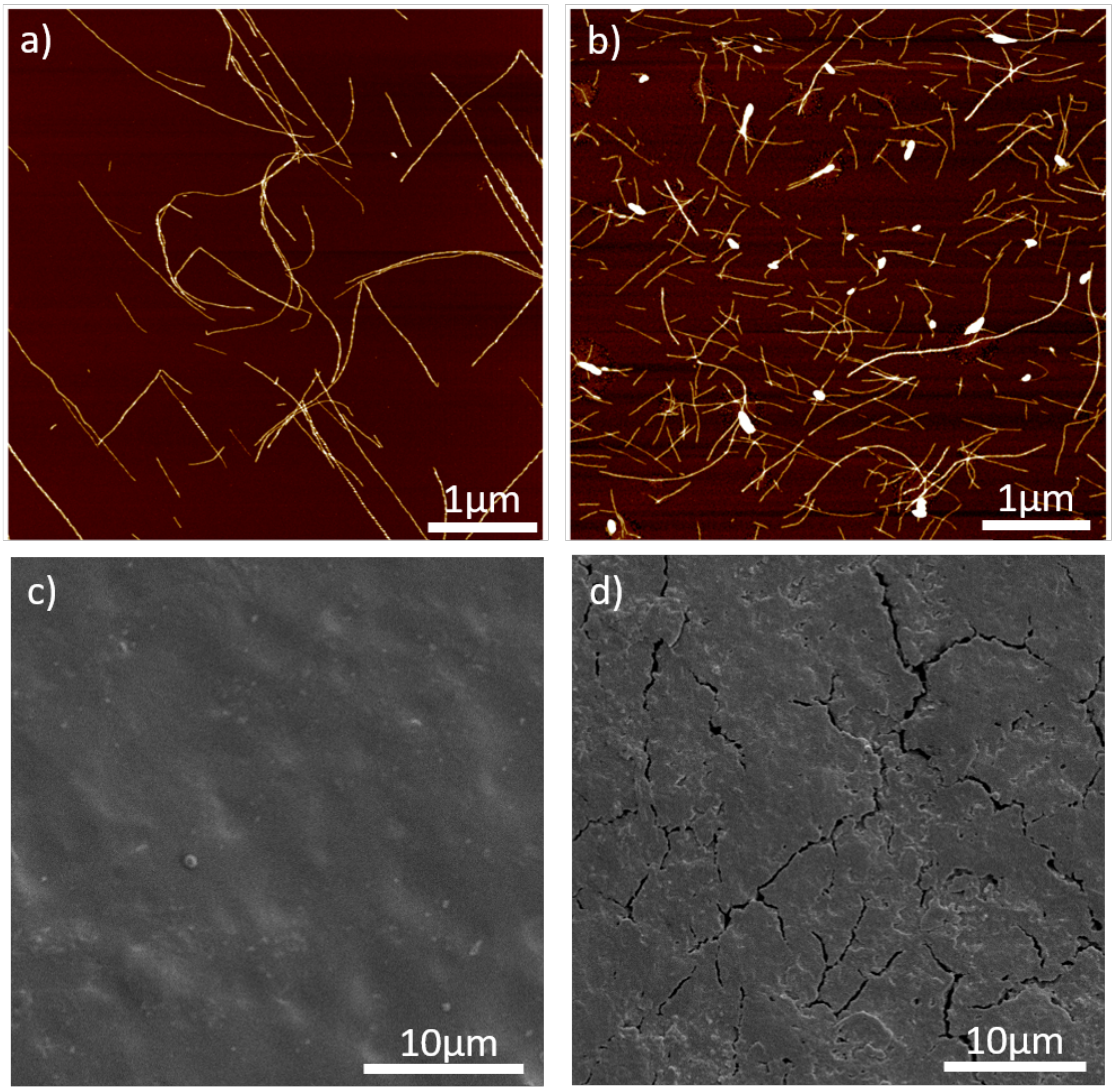

**Figure 2.** Characterization of amyloid fibrils and film surfaces. a) AFM of amyloid fibrils, b) AFM of the mixture of amyloid fibrils, Glycerol, and PVA, c) SEM of hybrid amyloid film top surface, and d) SEM of hybrid monomer film top surface.

**2.3. Surface contact angle**

One of the most significant weaknesses of bioplastics is their inherent low water stability, especially compared to petrol-based plastics[14,38–40]. Producing films that are not degraded by vapor or water exposure is essential for their real-world application. We performed static water contact angle analysis to determine the hydrophobicity of amyloid-based films **(Figure S1).** As depicted in **Figure 3a**, the contact angle varies highly depending on the composition and the chemical treatment. In particular, the film's contact angle formed by native WPI monomers is the smallest one, confirming the hydrophilic nature of this species. However, the contact angle measured on amyloid fibril-based films resulted in a more hydrophobic nature than the monomer-based ones. Most impoetantly, the chemical crosslinking treatment using FAS resulted in films with a surface characterized by a super-hydrophobic nature, with contact angles above 90°. The films containing CA, showed the lowest contact angle measured, even compared to protein monomers.

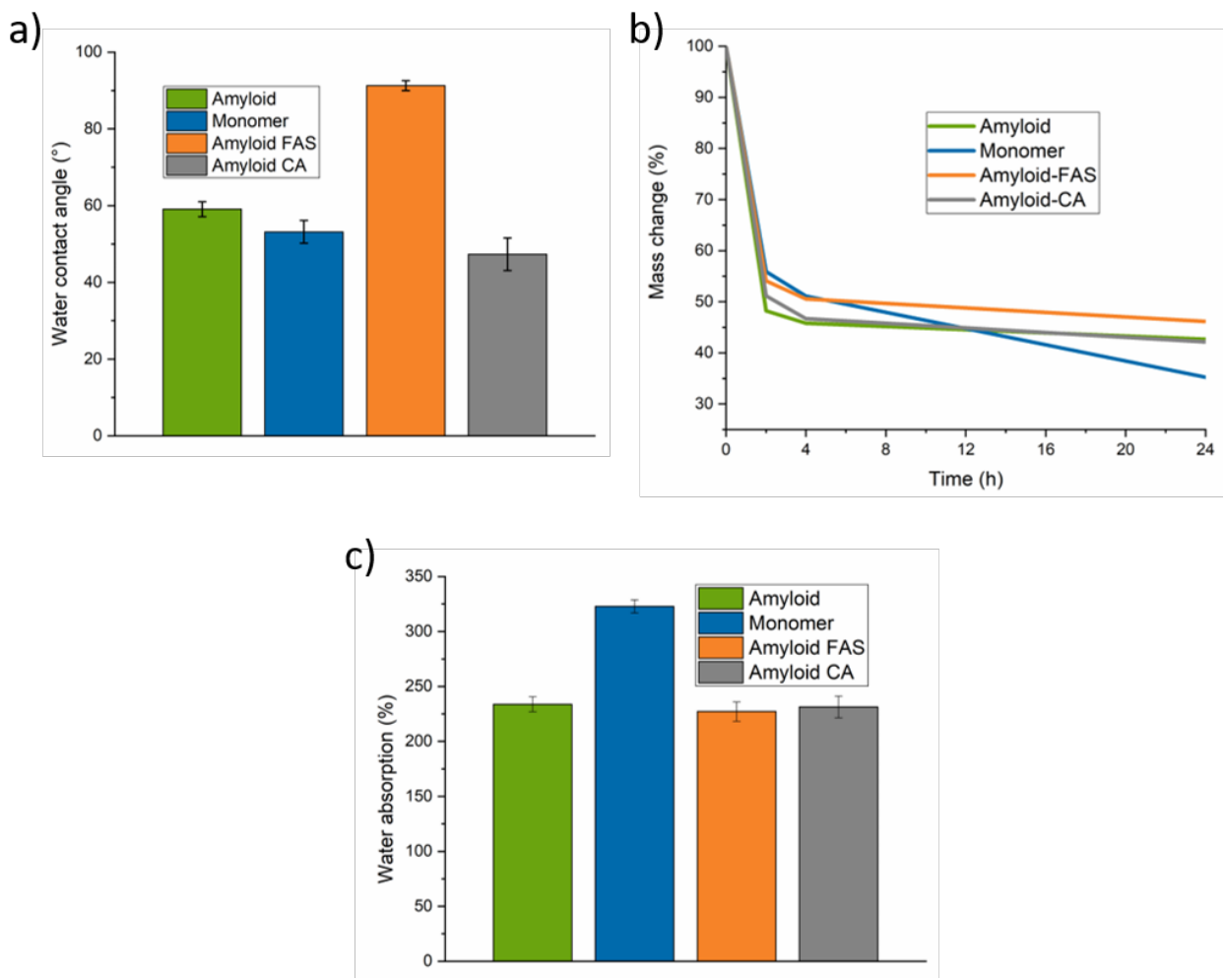

**Figure 3.** a) Water contact angle of the films after different crosslinking treatment. b) mass change, and c) water absorption.

## 2.4. Water stability

To further investigate the interaction of the films with water, we performed two additional experiments, the water absorption, and the mass release tests. As shown in **Figure 3b,** a substantial mass release of about 50% has been measured on all the films after 2 h of water immersion. However, after this first water immersion, the mass loss measured in films containing amyloid fibrils decreased drastically, leaving the mass essentially stable up to 24 h. The films' main component released in water is assumed to be primarily the plasticizer due to the high water solubility of this compound. This is further confirmed by the fact that the dried films appeared much more brittle than the same films before water immersion. At the same time, the films composed by WPI native monomers showed a continuous mass loss during all the time tested, up to complete film dissolution (not shown in the figure).

The water adsorption measurements resulted in very similar behavior in the films. As shown in **Figure 3c**, amyloid-based films absorbed around 225% of water after 24 h of immersion. The swelling, however, increased significantly in the films obtained by WPI monomers, reaching water adsorption values of around 325%. The low performances of monomer-based films are supposed to be related to the hydrophilic nature of the protein native state but also to the surface properties of these films. As already shown in Figures 2 and 3a, monomer-based films are characterized by a hydrophilic, inhomogeneous, and fractured surface, confirming the highest performances of amyloid fibrils compared to native monomers.

## 2.5. Mechanical properties

The mechanical properties of the hybrid whey amyloid, monomer, amyloid FAS-coated, and CA-contained films are depicted in **Figure 4**. As observed in stress-strain curves (**Figure 4a**), fibrillization, combination, and coating directly affect the mechanical properties of the films. The maximum stress at break, ultimate elongation, and toughness values are presented in **Figure 4b-d**, respectively (for Young's modulus, see **Figure S2**). The maximum stress of pure amyloid hybrid film is 17 MPa, similar to the value for monomer hybrid films. However, as shown in **Figure 4c**, the elongation of hybrid amyloid films is improved at least a factor two compared to hybrid monomer films. It was speculated that the better elongation of the amyloid hybrid films was attributed to the good alignment of nanofibril chains, allowing interfibrillar molecular rearrangement during deformation without undergoing fracture [41]. The excellent elongation to break of hybrid films is a very important property for food packaging applications. However, this property leads to a lower Young's modulus for hybrid amyloid fibrils films than monomer hybrid films (**Figure S2**).

Furthermore, it appeared that although either coating or hybridizing the films with FAS or CA, improves the maximum stress and Young's modulus, it decreases the film strain, resulted in rigid and less flexible films. This phenomenon is due to the crosslinking effect and limited mobility of whey amyloid fibrils chains after the addition of CA or coating with FAS [42]. Finally, the toughness of different hybrid films was measured as the area under the stress-strain curves, and the results are summarized in Figure 4d. As observed, all the hybrid amyloid-based films are characterized by higher toughness, which are up to 2 times higher than hybrid monomer ones.

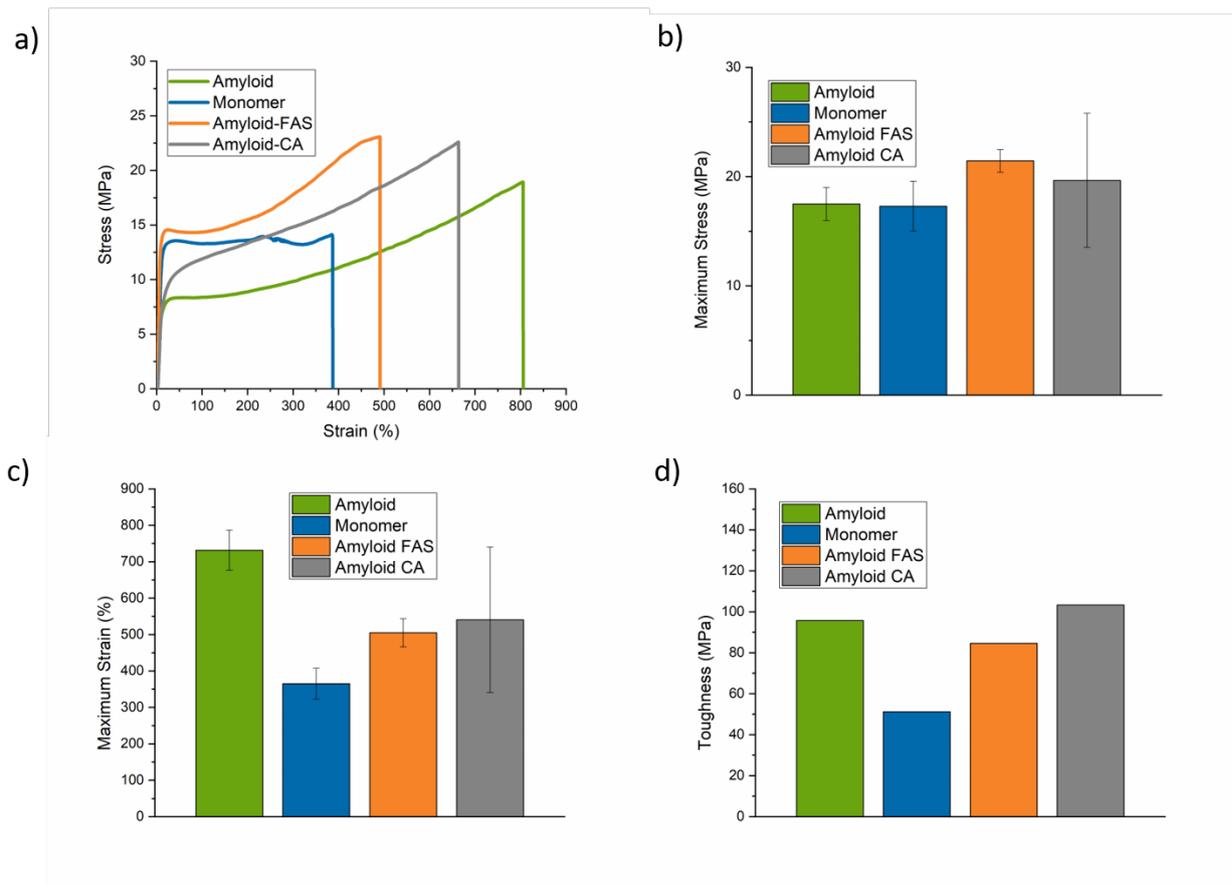

**Figure 4.** Mechanical properties of films. a) stress-strain plot, b) maximum stress, c) maximum strain, and d) toughness.

## 2.6. Optical properties

The optical properties of the films were analyzed using UV-vis spectroscopy, the resulting transmittance values in the visible spectrum (660 nm), and the UV range (280 nm) are summarized in **Table 2**. All the films tested are characterized by a high degree of transparency in the visible spectra with transmittance values above 95%. The films showed good UV-screening ability in the UV range with transmittance values dropping below 60%. This property is desirable in food-packaging materials since UV irradiation increases the oxidation rates of food, even at low temperatures [43]. At the same time, high transparency in the visible spectra is an important physical property, providing a see-through on the wrapped items. Hence, hybrid amyloid films features ideal optical properties for packaging applications.

**Table 1.** Optical properties of the prepared films.

|  | % Transmittance (660 nm) | % Transmittance (280 nm) |
|---|---|---|
| Hybrid amyloid | 95.9±0.4 | 57±7 |
| Hybrid monomer | 96.0±0.0 | 53±6 |
| Hybrid amyloid FAS | 97.4±0.1 | 58±2 |
| Hybrid amyloid CA | 97.2±0.1 | 56±2 |

## 2.7. Water vapor permeability

The barrier properties of the films against water vapor have been measured along 24 h, and the results are summarized in **Figure 5a**. Amyloid-based films showed the best performances, and the WVP resulted stable over the 24 h sampling. The hybrid films produced with WPI monomers showed WVP values higher than the hybrid amyloid films, and WVP drastically increased with time. Interestingly, the amyloid-based film showed extremely low WVP, and close to zero mass change was measured during the first 4 h of the test. On average, the hybrid monomer films showed WVP value $2.05 \times 10^{-6}$ g m m$^{-2}$ day$^{-1}$ Pa$^{-1}$, 1.5 times higher than the hybrid amyloid ones ($1.65 \times 10^{-6}$ g m m$^{-2}$ day$^{-1}$ Pa$^{-1}$).

## 2.8. Antioxidant activity

The antioxidant properties of the different hybrid films were analyzed with ABTS radical scavenging activity, and the results are presented in **Figure 5b**. The untreated hybrid amyloid films and FAS-treated amyloid films showed higher antioxidant performance when compared to hybrid WPI monomer films. Amyloid films treated with CA showed antioxidant properties such as those produced with WPI monomers. The antioxidant properties of the films are derived from the amino acids cysteine, tyrosine, tryptophan, and histidine, which are strong free radicals scavengers [44]. Importantly, this activity is enhanced for amyloid fibrils due to their significantly higher surface-to-volume ratio than protein monomers.

## 2.9. Food migration

To evaluate the performance of films as food packaging materials, the migration of components from the films to food was tested using Tenax®, a standard dry food simulant. As shown in **Figure 5c**, in hybrid amyloid films, a migration intensity of 3.6 mg/dm$^2$ is found, a value which is well below the limit (10 mg/dm$^2$) set by European Union legislation. Hybrid WPI monomer films, however, resulted in a migrations rate of 35.7 mg/dm$^2$, that is, well above the acceptable threshold.

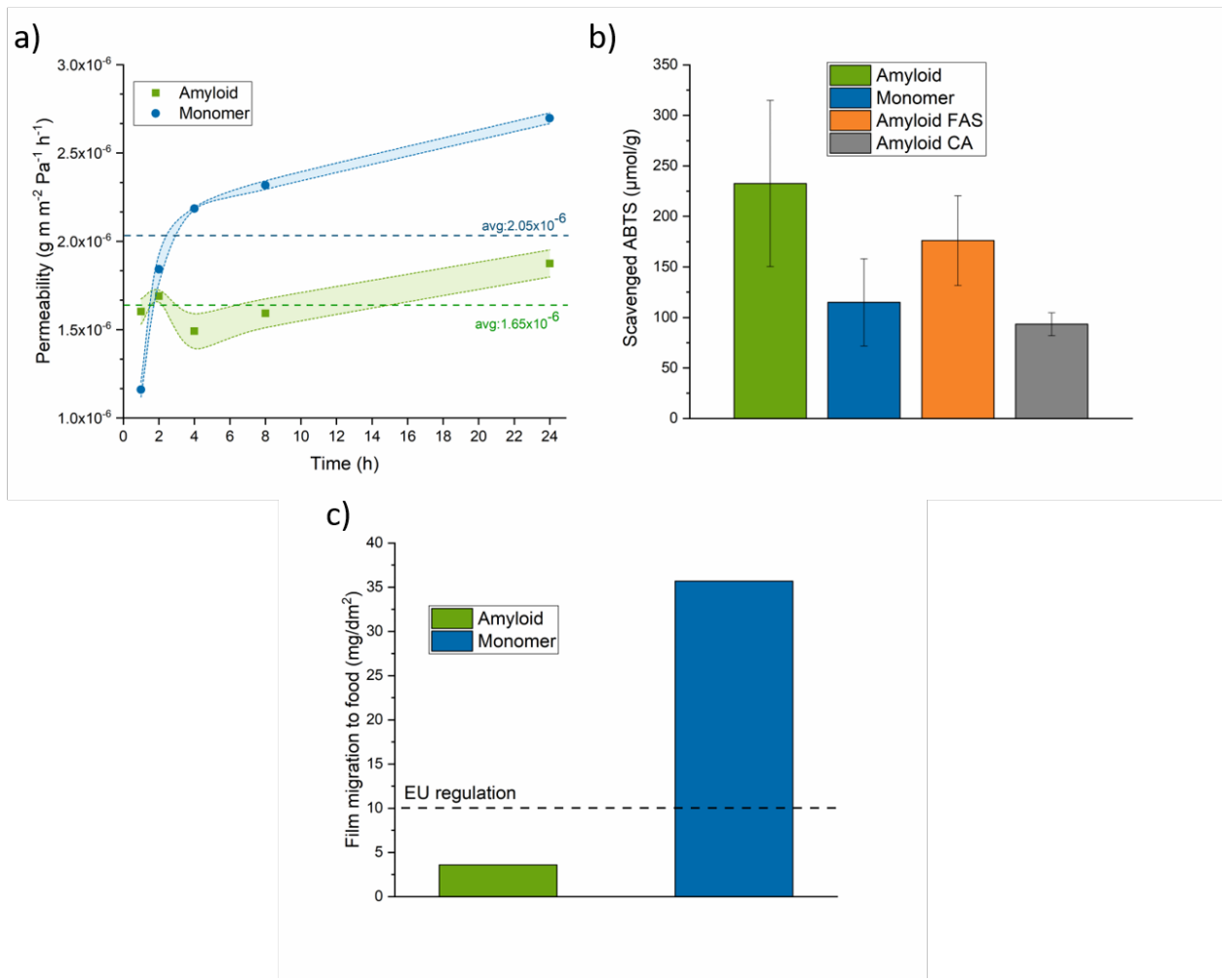

**Figure 5.** Film characterization. a) water vapor permeability. b) antioxidant activity. c) food migration analysis with Tenax®.

## 2.10. Generality, scalability, and benchmark to other plastics

To explore the generality of the approach for fabricating different hybrid amyloid fibrils, we replaced PVA with Methylcellulose (MC), methyl ether of cellulose. As observed in **Figure 6a**, the resulting HAm-MC film was smooth, flexible, and transparent. While both PVA and MC are biodegradable polymers, MC has the advantage of being from biosources and more sustainable. However, the water stability of native MC hybrid amyloid fibrils films was lower than the ones with PVA. Next, we demonstrate the scalability of the hybrid whey amyloid fibrils bioplastic by its production on a larger scale. As shown in **Figure 6b**, the bioplastic film with the size of 1.5 m×1.5 m can be produced successfully via a simple solution-casting process.

The mechanical properties of the hybrid amyloid fibrils films with PVA and MC were compared to different engineering, thermosets, rubbers and biodegradable plastics [45] (**Figure 6c**). Although both bioplastics in this work have similar elastic modulus, HAm-PVA has a higher elongation value compared to HAm-MC, reaching the value of 750%. This excellent strain value places the HAm-PVA film among the best elastic plastics ever produced. Even though the young modulus of bioplastics in this work was in the range of PTFE, PBS, PCL, LDPE, and EPDM, their elastic modulus was lower than the bioplastics such as PBS and PLA. However, precisely owing to such a high elastic modulus and a low elongation rate, it is not easy to make applicable bioplastics with PBS and PLA, and typically, they need to be blended with more flexible plastics for market applications[46]. This fact can be reflected in the toughness values shown in **Figure 6d**. Most of the biodegradable plastic suffer from low toughness values, resulting in products that are not ductile and are fragile. In case of amyloid based films, this property is significantly enhanced, resulting in toughness values several orders of magnitude superior to most of commonly used bioplastics.

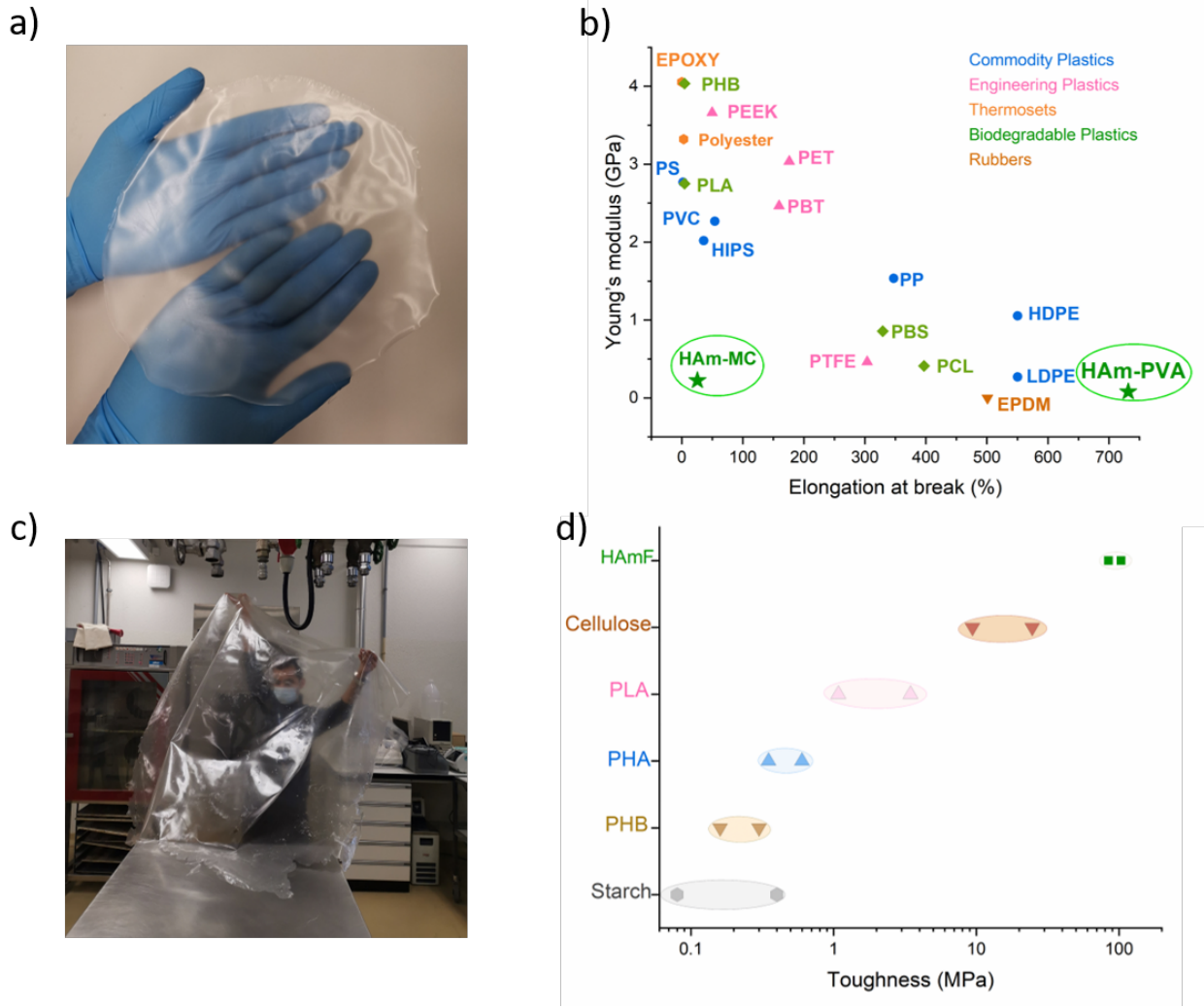

**Figure 6.** Generality, scalability, and comparison to other plastics. a) Visual appearance of the hybrid bioplastic based on amyloid fibrils and methylcellulose. b) scalability example of hybrid amyloid fibrils bioplastic. c) mechanical properties of different hybrid amyloid fibrils bioplastic and other plastics[45]. Panel d) shows the toughness values of bioplastics based on starch[47,48], PHB[49], PHA[50], PLA[49], cellulose[51] and hybrid amyloid films.

## 2.11. Life cycle assessment

LCA of HAm-PVA compared to PLA and PVF, including the exact amount of each category impact, is summarized in **Table 2**. Furthermore, the normalized environmental impact profile of production of the three plastics, comprising all 18 impacts, is presented in **Figure 7a**. Across all the impact categories, HAm-PVA bioplastic has the lowest environmental impact compared to PLA and PVF. This superiority is due to the HAm-PVA bioplastic greener and more straightforward production process. More importantly, the raw material for HAm-PVA bioplastic production is a waste from the dairy industry, which adds additional value to sustainability and circular economy. Combined to their excellent properties and degradability, the bioplastic presented in this work could lead to further environmental benefits not included in this LCA.

As observed, freshwater and marine ecotoxicities, human toxicity, and freshwater eutrophication have the highest relative contribution to the environmental impacts. The LCA results reveal that the main cause for the high values in these categories comes from the required energy for plastics production. To shed light on energy demand and correlate it with the other environmental impacts, we further performed CED analysis. The energy use breaks down into the categories of nonrenewable (fossil, nuclear, and biomass) and renewable (biomass, solar/wind/geothermal, and hydro) energy [52]. As shown in **Figure 7b**, PVF and PLA have 4 and 3 times higher energy consumption, respectively, than HAm-PVA bioplastic. **Figure 7c** shows the impact of the three plastics production on climate change. While the production of 1 kg of PVF and PLA approximately results in around 17 and 9 kg $CO_2$ eq, HAm-PVA bioplastic only emits about 4 kg $CO_2$ eq. Marine ecotoxicity is one of the most acute destructive effects of plastics on the environment since 10% of the plastic produced every year ends up in the oceans [53]. As observed in **Figure 7d**, the results for marine ecotoxicity showed the same trend as climate change impact, where HAm-PVA bioplastic showed minimal effect (0.02 kg 1,4-DB eq) compared to PLA (0.07 kg 1,4-DB eq), and PVF (0.16 kg 1,4-DB eq). Furthermore, as shown in **Figure 7e**, compared to the other two plastic, PLA has the highest impact on water resources, mainly due to water irrigation during crop production and high energy demand. Altogether, the LCA results demonstrate unambigously the superiority of HAm-PVA bioplastic in terms of sustainability and environmental compatibility compared to other typical plastics and bioplastics.

**Table 2.** LCA impact assessment based on ReCiPe Midpoint method.

| Impact category | Unit | HAm-PVA | PLA | PVF |
| --- | --- | --- | --- | --- |
| Climate change | kg $CO_2$ eq | 4.260198 | 9.220808 | 16.65613 |
| Ozone depletion | kg CFC-11 eq | 5.58E-07 | 4.76E-07 | 1.14E-06 |
| Terrestrial acidification | kg $SO_2$ eq | 0.020939 | 0.048895 | 0.078856 |
| Freshwater eutrophication | kg P eq | 0.001091 | 0.004544 | 0.007002 |
| Marine eutrophication | kg N eq | 0.001531 | 0.007004 | 0.002663 |
| Human toxicity | kg 1,4-DB eq | 0.934583 | 2.914094 | 6.327723 |
| Photochemical oxidant formation | kg NMVOC | 0.010556 | 0.027956 | 0.037809 |
| Particulate matter formation | kg PM10 eq | 0.007353 | 0.029297 | 0.042293 |
| Terrestrial ecotoxicity | kg 1,4-DB eq | 0.000881 | 0.008153 | 0.000708 |
| Freshwater ecotoxicity | kg 1,4-DB eq | 0.03041 | 0.079411 | 0.174013 |
| Marine ecotoxicity | kg 1,4-DB eq | 0.02145 | 0.070634 | 0.163069 |
| Ionising radiation | kBq U235 eq | 0.065584 | 1.161086 | 1.593614 |
| Agricultural land occupation | $m^2$a | 0.071638 | 1.518946 | 0.576349 |
| Urban land occupation | $m^2$a | 0.009264 | 0.102883 | 0.103839 |
| Natural land transformation | $m^2$ | 0.000828 | 0.001176 | 0.002372 |
| Water depletion | $m^3$ | 0.070116 | 0.322912 | 0.21498 |
| Metal depletion | kg Fe eq | 0.028109 | 0.115734 | 0.662357 |
| Fossil depletion | kg oil eq | 1.468327 | 2.324439 | 3.864043 |

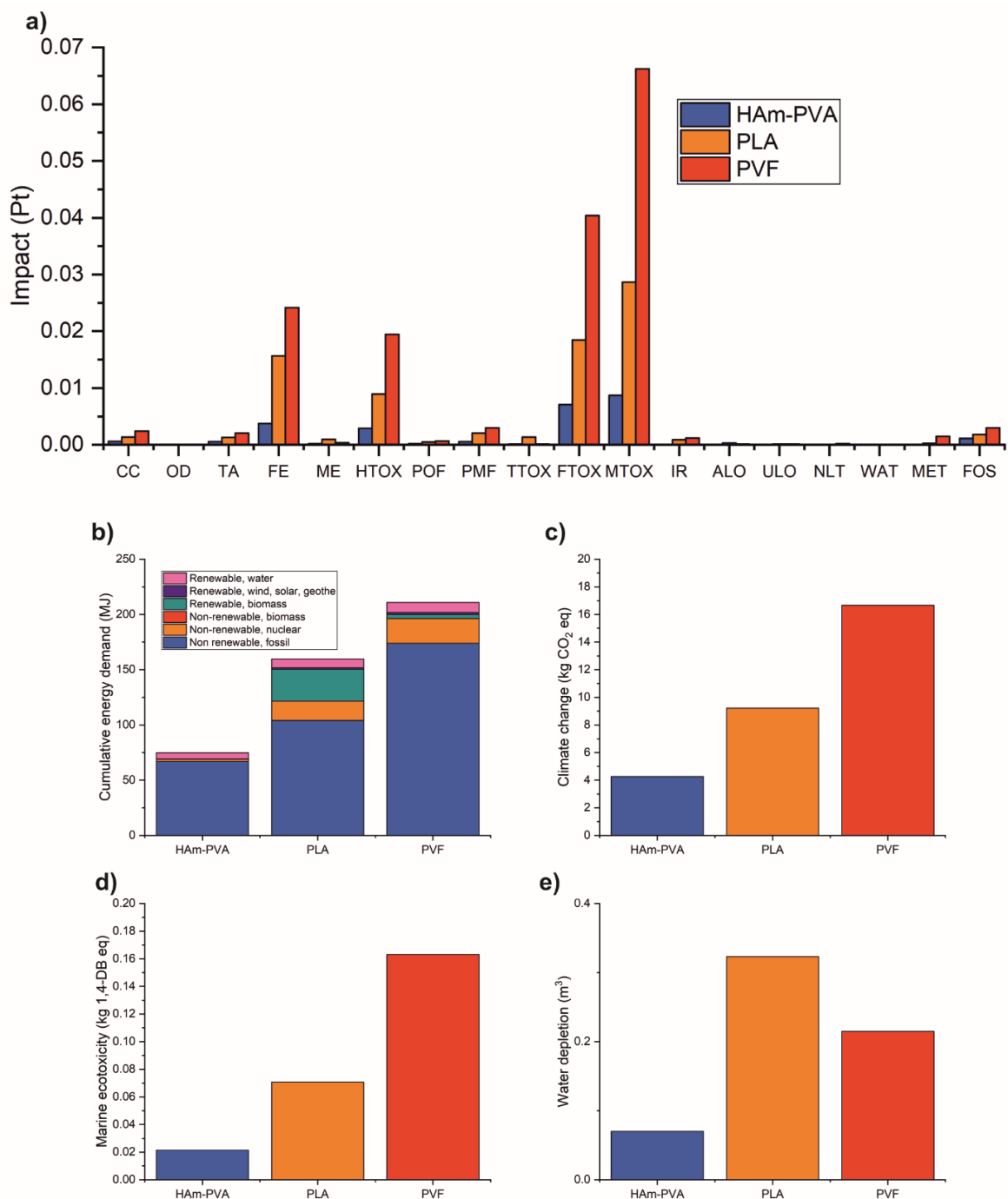

**Figure 7.** LCA of HAm-PVA compared to PLA and PVF. a) Normalized environmental impact profile of HAm-PVA compared to PLA and PVF, comprising all 18 impact categories of the ReciPe method (CC: climate change, OD: ozone depletion, TA: terrestrial acidification, FE: freshwater eutrophication, ME: marine eutrophication, HTOX: human toxicity, POF: photochemical oxidant formation, PMF: particulate matter formation, TTOX: terrestrial ecotoxicity, FTOX: freshwater ecotoxicity, MTOX: marine ecotoxicity, IR: ionizing radiation, ALO: agricultural land occupation, ULO: urban land occupation, NLT: natural land transformation, WAT: water depletion, MET: metal depletion, and FOS: fossil depletion). b) LCA results comparison based on cumulative energy demand impact. c) LCA results comparison based on climate change impact. d) LCA results comparison based on marine ecotoxicity. e) LCA results comparison based on water depletion impact.

## 3. Conclusion

In this work we introduced amyloid fibrils as a suitable building block for developing hybrid bioplastics. To this end, in situ fibrillization of whey monomer, selected here as model protein from food processing waste, took place in the presence of a plasticizer and a biodegradable plastic such as PVA and MC. The resultant films were transparent, robust, flexible, tough and exhibited acceptable water stabilities, as well as good barrier properties for food packaging applications. The films can be prepared from low-priced bio-based and biodegradable sources, highlighting their affordability and environment friendliness for a broad range of applications. Accordingly, LCA performed on these new bioplastics and two biodegradable polymers as benchmarks revealed a superior performance of the present bioplastics in all the normalized environmental impact indicators. Additionally, this new class of bioplastic adds value to circular economy by valorizing whey as a by-product of the dairy industry. These results demonstrated the potential of hybrid whey amyloid fibrils bioplastic as an efficient, sustainable, and inexpensive solution for alleviating the global plastics production and pollution issue.

## 4. Experimental Section/Methods

*Materials*: Whey protein isolate (WPI) was supplied from Fonterra, New Zealand. Polyvinyl alcohol (PVA, fully hydrolyzed, Mw approx. 200000) and Hydrochloric acid (36%) were purchased from Merck. Methyl cellulose (MC) (viscosity: 400 cP), Glycerol (≥99.5%), Citric acid (CA) (≥99.5%), 1H,1H,2H,2H-Perfluorooctyltriethoxysilane (FAS), 2,2′-azinobis(ethyl-2,3-dihydrobenzothiazoline-6-sulfonic acid) diammonium salt (ABTS), and Tenax® porous polymer adsorbent (60-80 mesh) were provided from Sigma Aldrich.

*Bioplastic film formation*: For fabricating hybrid amyloid fibril films, 4 g of WPI first were dissolved in 100 ml of water. Then the pH of the solution was adjusted to 2 and 3 g of Glycerol, as a plasticizer, were dispersed in the solution. For hybrid amyloid fibril films with PVA and MC, 3 g and 2 g of each compound, respectively were added to the solution. To convert the WPI monomers to amyloid fibrils and dissolving the biodegradable polymer, the solution was stirred and heated at 90 °C for 5 h. After the incubation, the solution was quenched and cast immediately on a petri dish to dry at room temperature. For the study of the effect of CA in film properties, 1.5 g were added to solutions before the fibrilization process. Moreover, for fabricating more hydrophobic films, they were placed in an ethanol solution containing FAS (0.5 wt %) for 1 h. Subsequently, the films were dried under room temperature to obtain the FAS-coated hybrid amyloid films. The details of composition and treatment are listed in **Table 3**.

**Table 3**. Bioplastic Films composition

| Bioplastics | Whey amyloid | Whey monomer | PVA | MC | Gly | FAS treatment |
|---|---|---|---|---|---|---|
| Hybrid amyloid | 40 wt.% | - | 30 wt.% | - | 30 wt.% | No |
| Hybrid monomer | - | 40 wt.% | 30 wt.% | - | 30 wt.% | No |
| Hybrid amyloid-FAS | 40 wt.% | - | 30 wt.% | - | 30 wt.% | Yes |
| Hybrid amyloid-CA | 40 wt.% | - | 30 wt.% | - | 30 wt.% | No |
| Hybrid amyloid-MC | 45 wt.% | - | - | 23 wt.% | 32 wt.% | No |

*Characterization*

*Atomic force microscopy*: Atomic force microscopy (AFM) and scanning electron microscopy (SEM) were used to characterize the morphologies of amyloid fibrils and their hybrid films, respectively. For AFM, the solutions were dried onto cleaved mica and analyzed by applying the tapping mode. A Hitachi SU5000 scanning electron microscope characterized the structure and properties of hybrid bioplastic films. Small pieces of films were attached to stubs with paste and sputter-coated with 5 nm of platinum/palladium under planetary rotational movement (Safematic, CCU- 10, Switzerland) before imaging.

*Mechanical properties*: The mechanical properties of films were evaluated by measuring tensile strength and elongation using a Z010 (Zwick) equipped with a 100 N load cell. The stress ($\sigma$)- strain (ε) curves were obtained at room temperature. The Young's modulus was calculated from the stress-strain curves.

*Water contact angle*: The water contact angle of the films was recorded by Nikon D300 digital camera at 25 °C and relative humidity of 50%. To shed light on the interaction of films with water, their weight loss, and water absorption after immersing in the water and at different time intervals were measured.

*Water absorption*: For the water absorption evaluation of films, approximately 150 mg of each film were immersed in water for 1 h, and the weight changes due to water absorption was measured before and after immersion.

*Antioxidant activity*: The antioxidant activity of film was determined by the spectrophotometric method described by Kusznierewicz et al.[54] Briefly, the stock solution of ABTS with the concentration of 7 mM was diluted with water to display the absorbance of 0.7 at 734 nm. Then 4.5 mL of ABTS solution was combined with a piece of film (10 mg). After 20 min reaction time, the film was removed, and the solution was transferred to a cuvette, and its absorbance was measured at 734 nm with the use of a UV−vis spectrophotometer (Cary 100, Agilent Technology). Finally, the amount of ABTS radicals scavenged by 1 g of the film was calculated based on the Beer-Lambert–Bouguer Law (equation 1):

$$Sc_{ABTS} = \frac{V_{ABTS} \times (A_0 - A_f) \times 1000}{\varepsilon \times l \times m} \quad (1)$$

where, $Sc_{ABTS}$ is the amount of scavenged ABTS (μmol), $V_{ABTS}$ is the volume of stock solution of ABTS added to the film (mL), $A_0$ is the absorbance of the initial ABTS solution; $A_f$ is the absorbance of the radical solution after reaction time; ε is the ABTS molar extinction coefficient (16,000 $M^{-1}$ $cm^{-1}$ at 734 nm), l is the optical path of the cuvette (1 cm) and m is the film mass (g) [54].

*Food-contact migration*: The food-contact migration properties of the bioplastic films were evaluated based on EU technical guidelines for compliance testing in the framework of the plastic FCM Regulation (EU) No 10/2011[55]. To assess the possible migration of molecules from the films to the food, Tenax® was used as a dry food simulant. In a clean glass petri dish, a square-shaped film with a dimension of 1 cm was placed between two layers of Tenax® powder (15 mg below and 15 mg above the sample) and stored in the oven for 2 h at 70 °C. The overall food migration was calculated by the mass difference of Tenax® before and after the treatment.

*Water vapor permeability*: The water vapor permeability (WVP) of the different hybrid films has been characterized using the modified cup method[56]. Briefly, glass vials with an inner diameter of 1.5 cm and a height of 6 cm were filled with 15 mL of water. The films were mounted on top of the vial using Parafilm to block air leakages. The weights of the vials have been measured after 1, 2, 4, 8, and 24 hours and the WVP estimated for each time step using the following formula:

$$\text{WVP} \left(g \cdot \frac{m}{m^2 \text{day Pa}}\right) = \frac{W \cdot t}{A \cdot T \cdot \Delta P} \tag{2}$$

Where W (g) is the weight decrease, t (m) is the film thickness, A (m$^2$) is the film surface area exposed to air, T (day) is the time, and $\Delta P$ is the difference in water vapor pressure between the inside and the outside of the vial (assumed 3173 Pa).

*Life cycle assessment*: The environmental impact of HAm-PVA bioplastic was compared via life cycle assessment (LCA) with one typical plastic film Polyvinyl fluoride (PVF), and one common bioplastic Polylactic acid (PLA). The LCA was an attributional and prospective LCA of an emerging product and performed according to the protocol of (ISO) 14040/44 standard [57,58]. The LCA assesses cradle-to-use life cycle impacts of producing 1 kg of these plastics. The Life cycle inventory (LCI) for all plastics is summarized in **Table S1**. For HAm-PVA bioplastic, the process data was provided by our laboratory experiments. For PLA, the inventory data for Ingeo® polylactide production technology from corn was used based on the LCA assessed by Suwanmanee et al[59]. Life cycle models were built using SimaPro v. 8.3 and based on Ecoinvent 3 database. ReCiPe midpoint (H) method was used to evaluate the impact of the LCI over a broad range of impact categories, 18 in this case. Additionally, cumulative energy demand (CED) was used for energy use calculations.

**Notes**
M.P and M.B. contributed equally to this work.

# Supporting Information

**Sustainable bioplastics from amyloid fibril-biodegradable polymer blends**

*Mohammad Peydayesh, Massimo Bagnani, and Raffaele Mezzenga\**

**Table S1.** Life cycle inventory data

| Life cycle inventory data | | | | | | |
|---|---|---|---|---|---|---|
| 1. HAm-PVA | | | | | | |
| Process Inputs | | | | | | |
| | Chemicals | Quantity | Units | SimaPro Process | Data base | Notes |
| Material | Whey | 800 | g | Liquid whey, from cheese production, at plant/NL Economic | EcoInvent | |
| | Water | 800 | ml | Water, process, surface | EcoInvent | |
| | PVA | 24 | g | Polyvinylchloride, bulk polymerised {GLO}\| market for \| Alloc Def, S | EcoInvent | Replacement for PVA |
| | Glycerol | 24 | g | Glycerin, at biodiesel plant/kg/RNA | EcoInvent | |
| | HCl | 4.4 | g | Hydrochloric acid, Mannheim process (30% HCl), at plant/RER Economic | EcoInvent | |
| | Equipment | Quantity | Units | SimaPro Process | Data base | Notes |
| Electricity | Heating and stirring plate | 0.966 | kW.h | Electricity, high voltage {ASCC}\| market for \| Alloc Def, S | EcoInvent | |
| Process Outputs | | | | | | |
| | Chemicals | Quantity | Units | SimaPro Process | Data base | Notes |
| | HAm-PVA | 168 | g | N/A | N/A | |
| 2. PLA | | | | | | |
| Process Inputs [1] | | | | | | |
| | Chemicals | Quantity | Units | SimaPro Process | Data base | Notes |
| Material | Maize grain | 1.54 | kg | Maize grain {GLO}\| market for \| Alloc Def, S | EcoInvent | |

| | Equipment | Quantity | Units | SimaPro Process | Data base | Notes |
|---|---|---|---|---|---|---|
| Electricity | Corn plantation, Diesel and chemicals | 6.4 | MJ | Energy, from diesel burned in machinery/RER Economic | EcoInvent | |
| | PLA pellet production | 10.6 | kW.h | Electricity, high voltage {GLO}| market group for | Alloc Def, S | EcoInvent | |

Process Outputs

| | Chemicals | Quantity | Units | SimaPro Process | Data base | Notes |
|---|---|---|---|---|---|---|
| | PLA | 1 | kg | N/A | N/A | |

3. PVF

Process Inputs

| | Chemicals | Quantity | Units | SimaPro Process | Data base | Notes |
|---|---|---|---|---|---|---|
| Material | PVF | 1 | kg | Polyvinylfluoride {GLO}| market for | Alloc Rec, S | EcoInvent | |

Process Outputs

| | Chemicals | Quantity | Units | SimaPro Process | Data base | Notes |
|---|---|---|---|---|---|---|
| | PVF | 1 | kg | Polyvinylfluoride {GLO}| market for | Alloc Rec, S | EcoInvent | |

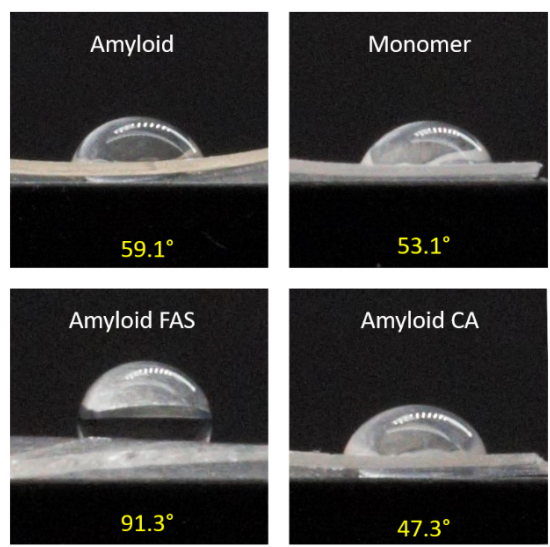

**Figure S1.** The images show photographs of water droplets standing on the different films.

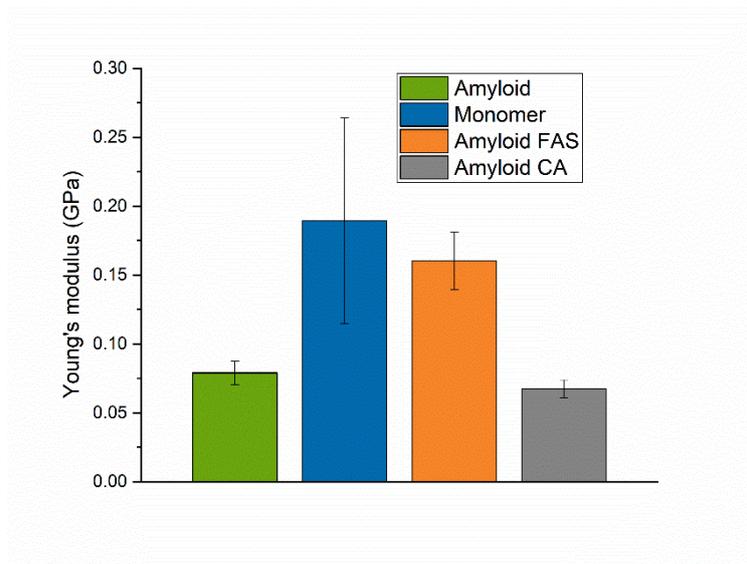

**Figure S2.** Young's modulus of the different films.

SI references